\documentclass{aa}

\usepackage{graphicx}
\usepackage{natbib}
\usepackage{amssymb}

\newcommand{\AaA}{A\&A}
\newcommand{\ApJS}{Astrophysical Journal Supplement Series}
\newcommand{\ApJ}{ApJ}
\newcommand{\Natur}{Nature}
\newcommand{\hi}{{\sc Hi} }
\newcommand{\hip}{{\sc Hi}}
\newcommand{\kmsp}{km s$^{-1}$}
\newcommand{\kms}{km s$^{-1}$ }
\newcommand{\cca}{cm$^{-2}$ }
\newcommand{\ccap}{cm$^{-2}$}
\newcommand{\ccu}{cm$^{-3}$ }
\newcommand{\ccup}{cm$^{-3}$}

\begin{document}

\title{Physical properties of a very diffuse HI structure at high Galactic latitude}

\author{Marc-Antoine Miville-Desch\^enes\inst{1,}\inst{2} \and Peter G. Martin\inst{2}}
\institute{Institut d'Astrophysique Spatiale, b\^atiment 121, Universit\'e Paris-XI, Orsay, 91405, France
\and Canadian Institute for Theoretical Astrophysics, 60 St-George st, Toronto, Ontario, M5S 3H8, Canada}
 
\offprints{Marc-Antoine Miville-Desch\^enes}
 \mail{mamd@ias.u-psud.fr}
\date{\today}

\titlerunning{A diffuse HI structure}
\authorrunning{Miville-Desch\^enes, M.-A. \& Martin, P.G.}

\abstract
{}
{The main goal of this analysis is to present a new method to estimate the physical properties of
diffuse cloud of atomic hydrogen observed at high Galactic latitude.} 
{This method, based on a comparison of the observations with fractional Brownian motion simulations, 
uses the statistical properties of the integrated emission, centroid velocity
and line width to constrain the physical properties of the 3D density and velocity fields,
as well as the average temperature of \hip. }
{
We applied this method to interpret 21 cm observations obtained with the Green Bank Telescope of a very
diffuse HI cloud at high Galactic latitude located in Firback North 1.
We first show that the observations cannot be reproduced solely by highly-turbulent CNM type gas
and that there is a significant contribution of thermal broadening to the line width observed.
To reproduce the profiles one needs to invoke two components with different
average temperature and filling factor.
We established that, in this very diffuse part of the ISM, 2/3 of the column density is made of WNM
and 1/3 of thermally unstable gas ($T\sim 2600$~K). The WNM gas is mildly supersonic ($<M>\sim1$)
and the unstable phase is definitely sub-sonic ($<M>\sim0.3$). The density contrast
(i.e., the standard deviation relative to the mean of density distribution)
of both components is close to 0.8. The filling factor of the WNM is 10 times
higher that of the unstable gas, which has a density structure closer
to what would be expected for CNM gas. This field contains a signature of CNM type gas
at a very low level ($N_H\sim 3\times 10^{19}$) which could have been formed by
a convergent flow of WNM gas. }
{}


\maketitle

\section{Introduction}

Neutral atomic hydrogen (\hip) is the most abundant phase of the interstellar medium (ISM) and 
it plays a key role in the cycle of interstellar matter from hot and diffuse plasma to the formation of stars. 
The 21 cm transition has been used since the 1950s \cite[]{ewen1951,muller1951} 
to study the properties of Galactic \hi but some of its basic properties are still unknown. 
The general picture of interstellar \hip, first modeled by \cite{field1969} 
(see also \cite{wolfire1995,wolfire2003}), is of spatially confined 
cold structures (CNM - Cold Neutral Medium, $T \approx 100$~K) 
surrounded by a more diffuse and warmer phase (WNM - Warm Neutral Medium, 
$T\approx 8000$~K). The heating and cooling mechanisms in the diffuse ISM
are such that these two phases can co-exist in pressure equilibrium. 
On the other hand the exact physical conditions in which the thermal
transition from WNM to CNM occurs -  most probably related to a local increase of the density and to the
temperature dependence of the cooling function - are still unclear.
Turbulence plays a dominant role in the kinematics and density structure of \hip:
both phases (CNM and WNM) have a self-similar density and velocity structure and
an energy spectrum compatible with a turbulent flow \cite[]{miville-deschenes2003c}.

With the advent of more sensitive and higher angular resolution observations, 
new challenges have come up (e.g., tiny scale atomic structures - TSAS -, 
self-absorption structures and thermally unstable gas).
An important advance came recently from the Millennium Arecibo Survey
conducted by \cite{heiles2003a,heiles2003}. These authors claimed that a significant fraction ($\sim 30$\%)
of the \hi is in a thermally unstable regime. This observational evidence seems to be
in accord with recent numerical simulations devoted to the study of the \hi thermal
instability \cite[]{audit2005}.

Here we contribute to this effort by analyzing 21 cm observations
of a very diffuse high-latitude region of the Galactic ISM.
The main advantage of studying the properties of weak \hi feature 
is that no modeling of radiative transfer is required to
interpret the brightness observed which can be converted directly in \hi column density.
Furthermore the 21 cm emission of such high latitude regions 
do not suffer from velocity crowding which facilitates the decomposition of the spectra. 
Our goal is to study the dynamical and thermo-dynamical properties of the neutral gas in
a diffuse region of the solar neighborhood.

Unlike previous analysis of 21 cm spectra we do not use a combination of emission and absorption 
observations to study the thermal and kinematical properties of the \hi
but instead we use the statistical properties of the emission on the plane of the sky
to infer the three-dimensional (3D) properties of the gas. In particular we exploit the
idea that variations of the value of the velocity centroid on the plane of the
sky can be used to constrain the amplitude of the turbulent motions on the line of sight
and therefore put limits on the average thermal broadening over the field.

In \S~\ref{sec_hi} we describe the 21 cm observations used for this analysis and
some quantities computed from them that will be used further to infer the \hi physical properties.
Then we describe (\S~\ref{section_simulation}) the method used to simulate realistic 21 cm observations 
using fractional Brownian motion and to test the hypothesis of an isothermal and turbulent \hip.
In \S~\ref{sec_two_components} we estimate the \hi physical properties in the hypothesis
of a two components model, and we conclude in \S~\ref{sec_conclusion}.

\section{The HI emission}

\label{sec_hi}

\subsection{GBT observations of the FN1 filament}

The present study focuses on a very diffuse \hi filament located
in the extra-galactic window known as Firback North 1 (FN1) \cite[]{dole2001} located
at Galactic coordinates (l=$84.0^\circ$, b=$45.1^\circ$).
The 21 cm \hi emission observations used in this analysis 
were obtained with the Green Bank Telescope (GBT). 
The size of the sub-field of FN1 observed is $1.1^\circ \times 1.1^\circ$, 
the half-power beam width is 9'.2 and the spectral
coverage spans the range from -169~\kms to 69~\kmsp, with a channel width
of $\delta u = 2.06$~\kms and a spectral resolution of 2.5~\kmsp. The noise level in a channel is 0.035 K.

\begin{figure}
\includegraphics[width=\linewidth, draft=false]{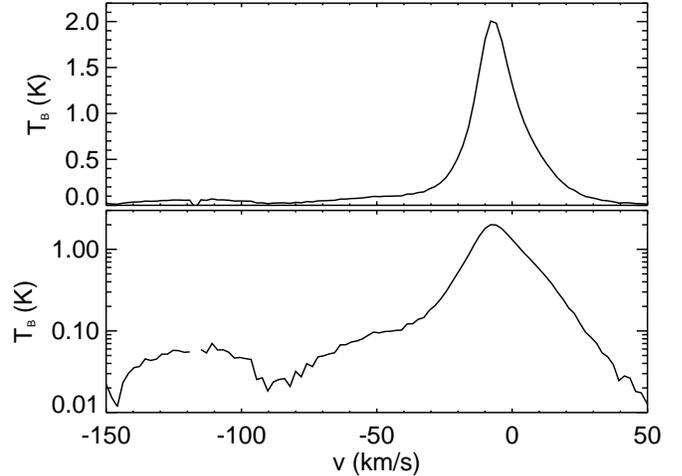}
\caption{\label{fig_avg_spectrum} GBT average 21 cm spectrum of the FN1 field
shown in linear-linear (top) and linear-log (bottom) scales. The spectrum
in linear-log scale helps to highlight the faint IVC ($u\sim -50$~\kmsp) and HVC ($u\sim-120$~\kmsp) components.}
\end{figure}

The 21 cm emission of this small patch of the sky is characterized by a single
spectral component at local velocities typical of \hi emission at high Galactic latitude, 
with a narrow core and extended wings (see Fig.~\ref{fig_avg_spectrum}). The GBT has
also detected a faint High-Velocity Cloud (HVC) centered at $u\sim-120$~\kmsp. The average \hi 
column density of the HVC over the field is $0.4\times 10^{19}$~\ccap with a maximum 
column density\footnote{The HVC component was computed by integrating over 
the channels between -168.5 and -96.4 \kmsp.} of $1.4\times 10^{19}$~\ccap.
The 21 cm spectra are also characterized by a weak ($\sim 0.1-0.2$~K) 
Intermediate Velocity Cloud (IVC) component centered at $u\sim-50$~\kms
(see Fig.~\ref{fig_avg_spectrum}-bottom).

\begin{figure*}
\includegraphics[width=\linewidth, draft=false]{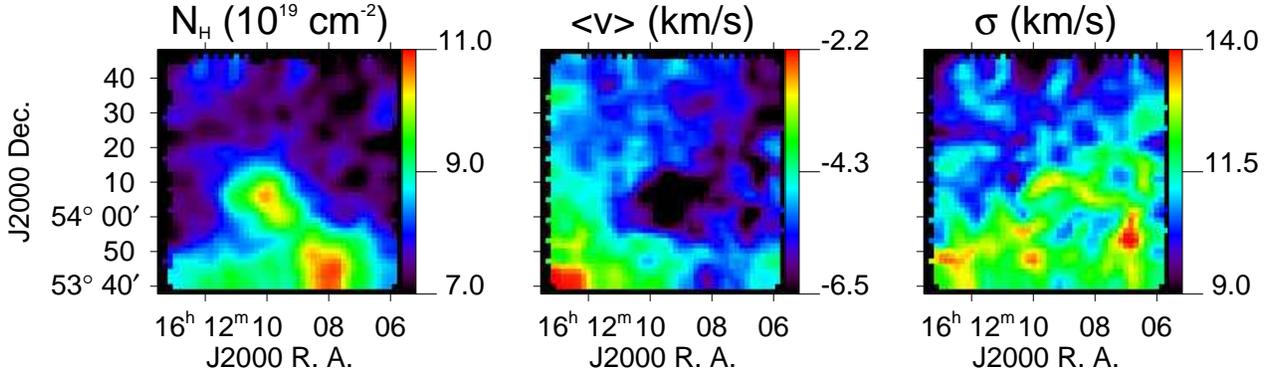}
\caption{\label{fig_resDecomp_OneCpn} \hi column density (left),
centroid velocity (middle) and velocity dispersion (right column) of the
local \hi gas in FN1.}
\end{figure*}

\subsection{Properties of the \hi emission spectra}

The 21 cm emission profiles observed are shaped by density, velocity
and thermal fluctuations of interstellar neutral hydrogen on the line of sight.
In this section we define quantities that can be deduced from the observations
and that will be used further to constrain the physical properties
of the region observed.

\subsubsection{Brightness temperature of an optically thin line}

In the optically thin limit where 
\hi self-absorption is negligible \cite[]{spitzer1978}, the brightness
temperature (in K) observed at position $(x,y)$ on the sky and in the velocity range
between $u$ and $u+du$ is:
\begin{eqnarray}
\label{icube}
T_B(x,y,u)\, du & = & B(x,y) \otimes \frac{1}{1.823\times 10^{18}\Delta} \int_0^H n(x',y',z) \, dz \nonumber \\
& & \times \, \frac{1}{\sqrt{2\pi}} \exp(-\frac{(u-v(x',y',z))^2}{2\Delta^2}) \, du.
\end{eqnarray}
where $n(x,y,z)$ and $v(x,y,z)$ are the  three-dimensional
density and velocity fields.
The term $\Delta^2 = k\,T(x,y,z)/m + \sigma_{psf}^2$ represents the thermal broadening of the 21 cm line
and the instrumental broadening\footnote{here we consider
that the spectral response follows a Gaussian function}; 
$T(x,y,z)$ is the 3D temperature field, $B(x,y)$ is the instrumental beam,
$H$ is the depth of the line of sight, $m$ is the hydrogen atom mass and $k$ is the 
Boltzmann constant\footnote{Here we make the assumption that the depth of the cloud is small compared
to the distance to the cloud, which allows us to do the integral in Cartesian
coordinates and not take into account the fact that a beam sees structures
of increasing physical size with distance.}.
Even when radiative transfer effects are negligible, like for the diffuse
filament observed here, the 21 cm emission depends in a complex manner on 
the density, velocity and temperature distribution on the line of sight.
In addition several studies showed (e.g., \cite{miville-deschenes2003c}) 
that the \hi density and velocity 
distributions are self-similar, which reflect the impact
of turbulence on the interstellar medium dynamics. 

One of the challenges of the analysis of 21 cm emission is to 
disentangle density, velocity and temperature 
fluctuations, using quantities averaged along the line of sight.
To constrain the properties of the density, velocity and
temperature of the gas in 3D we use the integrated emission map $N_H(x,y)$,
the centroid velocity map $C(x,y)$ and the velocity dispersion
map $\sigma(x,y)$. 

\subsubsection{Integrated emission}

In the optically thin limit, the integrated emission is proportional to
the column density (in $cm^{-2}$), which is simply
the integral of the density fluctuations on the line of sight:
\begin{eqnarray}
\label{eq_column_density}
N_H(x,y) & = & 1.823\times 10^{18} \sum_u T_B(x,y,u) \delta u\\
& = & B(x,y) \otimes \int_0^H n(x',y',z) \, dz.
\end{eqnarray} 

\subsubsection{Velocity centroid}

The velocity centroid map is the average of the velocity as 
weighted by the brightness temperature at each position on the sky:
\begin{equation}
\label{eq_centroid1}
C(x,y)=\frac{\sum_u u~T_B(x,y,u)}{\sum_u T_B(x,y,u)}.
\end{equation}
In the optically thin limit, the velocity centroid map is equal to
the following expression:
\begin{equation}
\label{eq_centroid2}
C(x,y) =  \frac{B(x,y) \otimes \int_0^H n(x',y',z) \, v(x',y',z) \, dz}{B(x,y) \otimes \int_0^H n(x',y',z) \, dz}.
\end{equation}
The velocity centroid map is thus the density-weighted average velocity
on each line of sight. It depends non-linearly on the density and
velocity distribution in 3D, but not on the gas temperature.

\begin{table*}
\begin{center}
\caption{\label{table_gbt} Statistical properties of the local \hi emission.}
\begin{tabular}{lccccc}\hline
& $<N_H>$ & $\sigma_N$ & $<C>$ & $\sigma_C$ & $<\sigma>$\\
& ($10^{19}$ cm$^{-2}$) & ($10^{19}$ cm$^{-2}$) & (\kmsp) & (\kmsp) & (\kmsp) \\\hline
Total & $8.1\pm0.1$ & $0.98\pm0.02$ & $-5.2\pm0.2$ & $0.9\pm0.1$ & $11.3\pm0.6$ \\
Narrow & $2.48\pm0.01$ & $0.77\pm0.03$ & $-7.27\pm0.01$ & $0.75\pm0.02$ & $4.96\pm0.01$\\
Wide & $5.89\pm0.01$ & $0.5\pm0.1$ & $-2.96\pm0.01$ & $1.66\pm0.06$ & $13.44\pm0.01$\\ \hline
\end{tabular}
\end{center}
The {\em Total} row gives the statistical properties computed on the total emission (excluding the HVC). 
The uncertainties on the {\em Total} values represents their variance when computed by selecting data points 
with a 1 and 3 times the noise level threshold.
The {\em Narrow} and {\em Wide} rows give the statistical properties of the
two components resulting from the Gaussian decomposition (see \S~\ref{sec_gaussian_decomposition}). 
In these cases the uncertainties
were computed using the uncertainties on the parameters of the fit (see text for details).
\end{table*}

\subsubsection{Velocity dispersion}

The second velocity moment, the velocity dispersion map, is the velocity dispersion
on the line of sight. It is computed the following way:
\begin{equation}
\label{eq_dispersion1}
\sigma^2(x,y) = \frac{\sum_u u^2 \, T_B(x,y,u) \, \delta u}{\sum_u T_B(x,y,u) \delta u} - C^2(x,y).
\end{equation}
This quantity is more affected by instrumental noise but it allows one to
estimate the gas temperature. It is related to the 3D quantities
by the following equation:
\begin{equation}
\sigma^2(x,y) = \sigma^2_{turb}(x,y) + \sigma^2_{psf}(x,y) + \sigma^2_{therm}(x,y)
\end{equation}
where  $\sigma_{psf}$ is the instrumental broadening due to the finite spectral resolution,
\begin{equation}
\label{eq_sigmaturb}
\sigma^2_{turb}(x,y) = \frac{B(x,y) \otimes \int_0^H n(x',y',z) \, v^2(x',y',z)\, dz}{B(x,y) \otimes \int_0^H n(x',y',z) \, dz} - C^2(x,y)
\end{equation}
is the broadening from turbulent motions and
\begin{equation}
\sigma^2_{therm}(x,y) =\frac{k}{m}\frac{B(x,y) \otimes \int_0^H n(x',y',z) \, T(x',y',z) \, dz}{B(x,y) \otimes \int_0^H n(x',y',z) \, dz}
\end{equation}
is the thermal broadening.

\subsection{Statistical properties deduced in FN1}

The \hi integrated emission, centroid velocity and velocity dispersion maps of the FN1 sub-field, 
computed respectively using Eq.~\ref{eq_column_density}, Eq.~\ref{eq_centroid1} and Eq.~\ref{eq_dispersion1},
are shown in Fig.~\ref{fig_resDecomp_OneCpn}. 
These three maps were computed by using only data points with a value higher than 2 times the noise level 
in a single channel.
The average and standard deviations of these three maps\footnote{The standard deviation of $\sigma(x,y)$ 
is not given here and not used in this
analysis as it is more affected by noise} 
are given in the first row of Table~\ref{table_gbt} (labeled ``Total'').
To estimate the uncertainties listed we have computed
the same three maps by selecting data points with a value higher than 1 time and 3 times the noise level.

\subsection{Discussion}

Without any 21 cm absorption measurements in that field,
our idea was to use the statistical properties of the projected quantities
together with the spectral shape of the 21 cm line (Fig.~\ref{fig_avg_spectrum})
as the basis for some constraints on the level of turbulence.
More specifically we wanted to estimate the relative contributions
of the temperature and turbulence fluctuations to the line profile and to study 
if the observed profiles can be the result of relatively cool gas with strong turbulent motions. 
As stated previously the statistical quantities given in the first row of Table~\ref{table_gbt} represent
a mixture of density, velocity and temperature fluctuations on the line-of-sight.
They depend on the statistical properties
of the 3D density, velocity and temperature fields, and of course on the depth of the line-of-sight.

\begin{figure}
\includegraphics[width=\linewidth, draft=false]{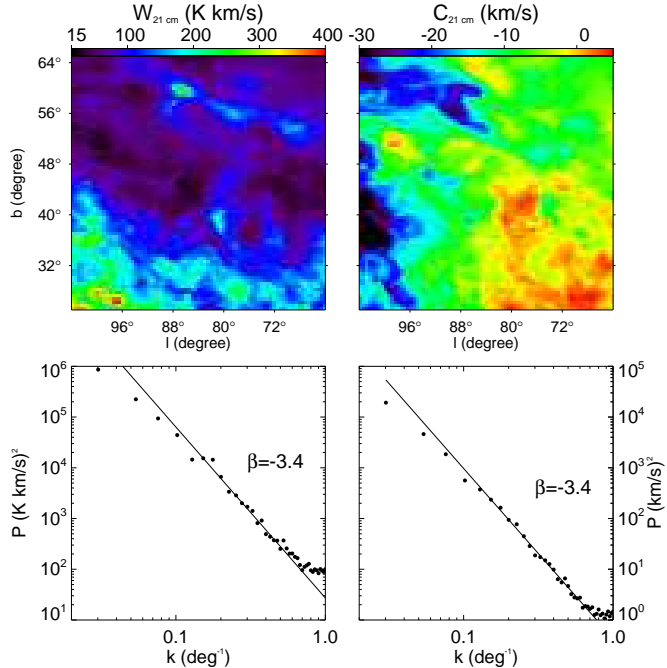}
\caption{\label{fig_powerspectra_leiden} Power spectrum analysis of
the \hi emission in the FN1 region. {\bf Top:} 21 cm integrated emission
(left) and centroid velocity map (right) obtained with the Leiden/Dwingeloo
data \cite[]{burton1994}. The two maps were computed using velocity channels
from -86.7~\kms to 47.2~\kms to avoid contamination by high-velocity clouds.
{\bf Bottom:} Power spectrum of the 
above maps with corresponding power law fit. Both power spectra
were divided by the instrumental response
of the Leiden antenna (beam of 30 arcmin FWHM). }
\end{figure}

\begin{figure}
\includegraphics[width=\linewidth, draft=false]{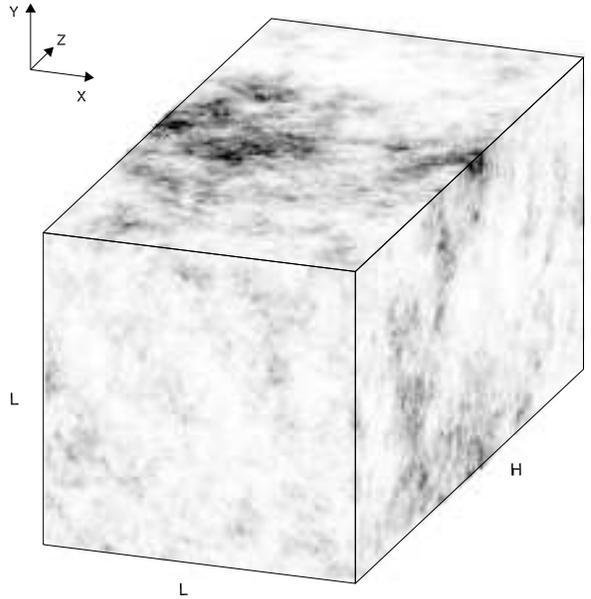}
\caption{\label{fig_cubesimu} Simulation of a $65\times 65 \times 128$ 
density field with $\sigma(n)/<n>$=1 and $\gamma_n=-3.4$. The figure shows the
density fluctuations at surface of three sides of the cube.}
\end{figure}

\section{Simulation of the GBT observation}

\begin{figure}
\includegraphics[width=\linewidth, draft=false]{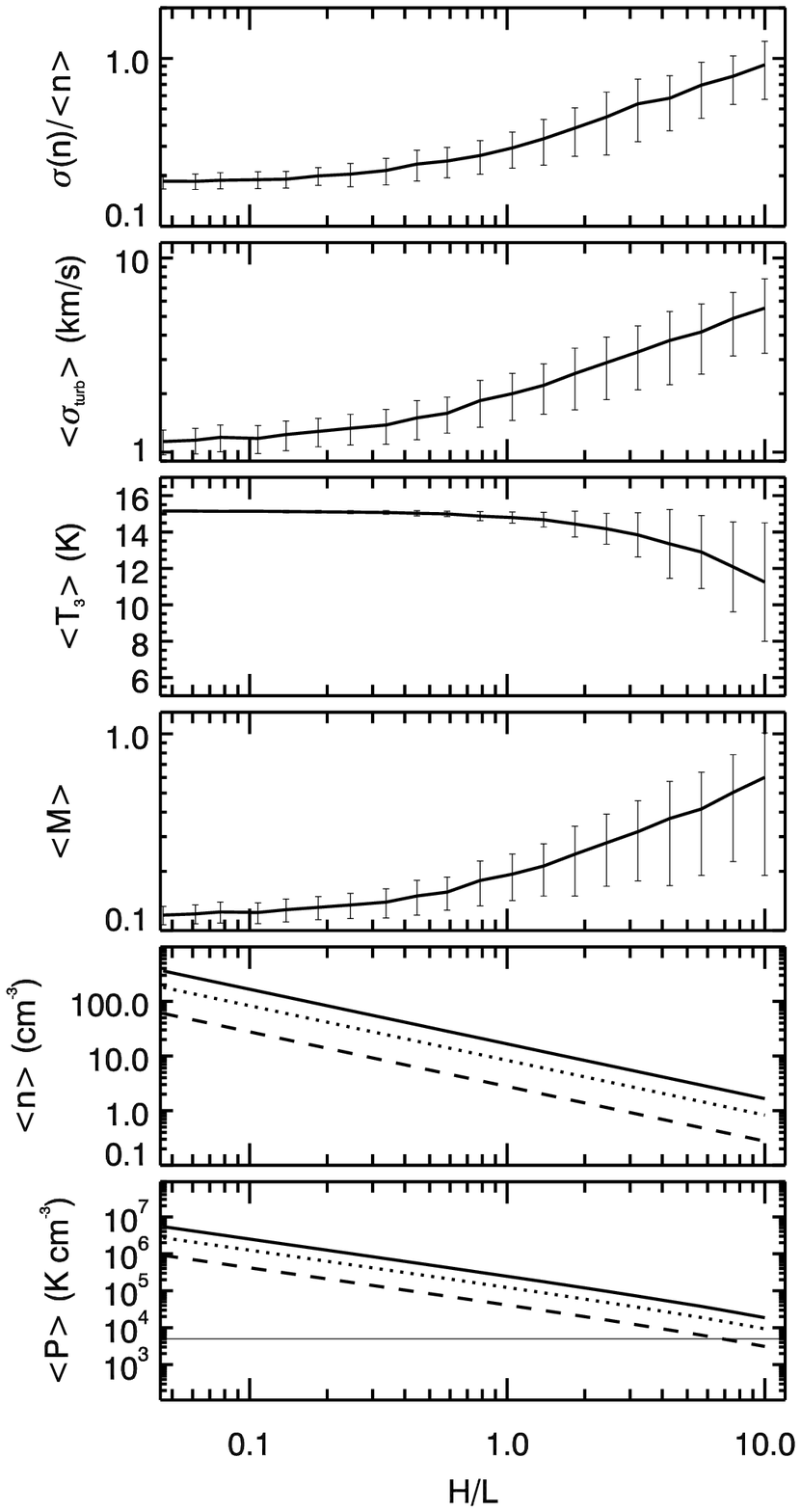}
\caption{\label{fig_resSimulOneCpn} One component model : physical parameters (density contrast $\sigma(n)/<n>$, 
average turbulent broadening $<\sigma_{turb}>$, average temperature $<T_3>=T/1000$, average Mach number $<M>$)
as a function of the axis ratio $H/L$. Also given are the average density $<n>$ and average pressure $<P>$
for distances to the cloud of 75 pc (solid), 150 pc (dotted) and 450 pc (dashed). }
\end{figure}

\label{section_simulation}

In this section we describe how we produced simulated GBT observations to 
constrain the properties of the 3D density $n$ and velocity $v$ fields
as well as the average kinetic temperature of the gas. 
One solution to simulate GBT observations would be to use 
hydrodynamics or magneto-hydrodynamics simulations
of the \hi that would include the thermal instability (e.g., \cite{audit2005}).
Such simulations give a close representation of the
actual physics. On the other hand they do not provide a direct handle on
the power spectrum of the density and velocity, nor on the average
density and velocity, or on the axis ratio of the cloud (i.e., depth over extent
on the sky). Finally, doing such simulations is computationally intensive and it is not
practical to make a large number of them to estimate statistical variance
of the results.

\begin{figure*}
\includegraphics[width=\linewidth, draft=false]{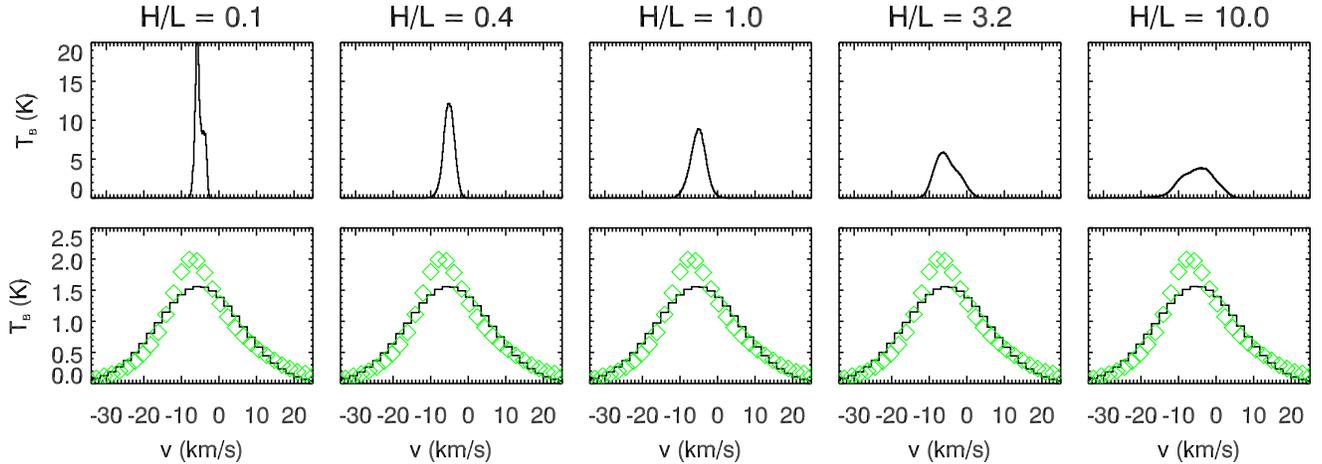}
\caption{\label{fig_simu_one_component} Comparison of simulated and observed spectra.
{\bf Top: } Typical simulated spectra, averaged over the $1.1^\circ \times 1.1^\circ$ field,
for 5 values of the axis ratio $H/L$. These spectra represent only the turbulent part of the
broadening. {\bf Bottom: } Observed average GBT spectrum (diamonds) compared to the
average simulated spectra in which thermal and instrumental broadening
were added to the turbulence to reproduce the observed $<\sigma>$.}
\end{figure*}

To have explicit control on the statistical properties of the fields and
the axis ratio of the cloud, and to be able to make a large number of realizations,
we chose to simulate 3D velocity and density fields
using fractional Brownian motion (fBm), which are Gaussian random fields 
with a given power spectrum \cite[]{miville-deschenes2003b}.
To simulate the GBT observation
we made the assumptions that we are observing a single \hi cloud
on the line of sight, that its 3D density and velocity fields
are self-similar and characterized by power law power spectra
and that the gas structure and kinematics are statistically isotropic in three dimensions.
In this framework the statistical properties (mean and variance) of the observed 
quantities ($N_H(x,y)$, $C(x,y)$ and $\sigma(x,y)$) 
will depend on a relatively limited number of parameters:
the index of the density and velocity power spectra
($\gamma_n$ and $\gamma_v$), the 3D average density $<n>$, the
3D density dispersion $\sigma_n$,  the 3D average velocity $<v>$, the 3D velocity 
dispersion $\sigma_v$, the average gas temperature $<T>$,
the cloud depth along the line of sight $H$ and the distance to the cloud $D$.
In fact we reached the conclusion that for given value
of $\gamma_n$, and to reproduce given $<N_H>$ and $\sigma_N$, there is a close one to one 
relation between the required ratio $\sigma_n/<n>$ and 
the cloud axis ratio $H/L$, where $L$ is the physical size of the field on the plane of the sky, 
that is independent of the distance $D$ to the cloud.
The model $\sigma_v$ and observed $\sigma$ are also independent of distance and 
only depend on the density contrast $\sigma_n/<n>$.
It is thus convenient to produce simulations for a constant distance $D$ (we adopted 100 pc)
but with a variable $H/L$.

\subsection{Power spectrum of the \hi emission}

\label{section_ps_leiden}

One important ingredient of our analysis is the 
power spectrum of the density and velocity fluctuations in 3D.
\cite{miville-deschenes2003b} showed that the power spectrum of 21 cm
integrated emission and of the centroid velocity fields can be used to estimate 
the spectral indexes $\gamma_n$ and $\gamma_v$ of the 3D density and velocity 
fields\footnote{As shown by \cite{brunt2004} and \cite{esquivel2005} this is only true for 
low contrast density fields ($\sigma_n/<n> \leq 1$).}.
\cite{miville-deschenes2003c}
used this property to estimate the power spectrum of velocity fluctuations
in 3D of a high-latitude cloud.
Such power spectra cannot be computed solely with the GBT observations which
have  a very limited spatial dynamic range (from 9.2' to 1.1$^\circ$). Instead we
used the 21 cm Leiden/Dwingeloo data \cite[]{burton1994} of a
$40^\circ\times40^\circ$ region centered on the GBT field. The integrated emission
and centroid velocity maps extracted from these data are shown in 
Fig.~\ref{fig_powerspectra_leiden} together with their respective power spectrum.
We have taken great care here to select velocity channels with emission
from the local gas only. The power spectra of the integrated emission and
centroid velocity are both well fitted by a slope of $-3.4\pm0.2$ on scales
smaller than $5^\circ$. {At larger scales the power spectra flatten slightly
which could be partly due to an effect of the tangent plane projection.
It could also be an indication that we have reached the angular scales on
the sky that correspond to the depth of the medium (i.e., the transition from
2D to 3D) \cite[]{miville-deschenes2003b,elmegreen2001,ingalls2004}. }

It is interesting to note at this point that the spectral indices measured here are
compatible with what has been measured in another high-latitude field 
by \cite{miville-deschenes2003c} also using 21 cm observations.
In the GBT simulations we consider that the power spectrum indices of the 
3D density and velocity fields are $\gamma_n=\gamma_v=-3.4$ following 
what is observed at larger scales.

\subsection{Simulations of density and velocity fields}

We made simulations of $n(x,y,z)$ and $v(x,y,z)$ using fBm \cite[]{miville-deschenes2003b}
with $\gamma_n=\gamma_v=-3.4$.
We performed 150 simulations each for 20 different values of the $H/L$ ratio 
(from 0.05 to 10, equally spaced logarithmically).
{The size of our cubes were from $65\times 65\times 3$ pixels (for $H/L=0.05$) 
to $65 \times 65 \times 650$ pixels (for $H/L=10$). 
To minimize the effect of periodicity of fBm objects 
each cube of size $65 \times 65 \times N$ was in fact extracted at a random position
from a larger fBm cube of $1024^3$ pixels. 
For each of these 3000 simulations the $n(x,y,z)$) and $v(x,y,z)$) fields were constructed independently
with no correlation between them\footnote{\cite{miville-deschenes2003b,brunt2004} showed that
the level of correlation between the density and velocity does not modify
significantly the statistics of projected quantities.}. 

Each pair of ($n$, $v$) cubes were then combined to produce a brightness temperature cube $T_B(x,y,u)$
(using Eq.~\ref{icube}) on the grid of the GBT observation 
($48 \times 48$ pixels times 116 channels of width $\delta u=2.06$~\kmsp)
and taking into account the spatial and spectral instrumental response of the GBT.
The pixel size in our original $64\times64\times N$ cube is 1.1' which has to be compared
with the GBT beam of 9'. We have tested that this was enough spatial resolution to take
into account fluctuations at scales smaller than the beam.

To normalize the cubes $n$ and $v$ 
we computed the column density ($N_H(x,y)$) and centroid velocity ($C(x,y)$) maps
from the simulated $T_B(x,y,u)$ using equations~\ref{eq_column_density} and \ref{eq_centroid1}. 
For the velocity, the normalized field is simply
\begin{equation}
v' = A v + B.
\end{equation}
We find $A$ and $B$ such that our simulation reproduces the observed values
$<C>$ and $\sigma_c$.

For the density field the normalization process is more complicated because it has to be positive everywhere.
Being Gaussian, fBm fields are limited to contrast lower than 1/3
as 99\% of the fluctuations are within 3$\sigma$ of the mean. Interstellar density fields
are expected to have higher contrast than 1/3 which calls for a modification of the
usual fBm. To produce positive and higher-contrast fBm we use the following method
\begin{equation}
\label{eq_modif_fbm}
n' = \alpha (n-n_{min})^\beta
\end{equation}
where $n_{min}$ is the minimum value of $n$.
We estimate $\alpha$ and $\beta$ so that the simulated column density map has the same  $<N_H>$ and $\sigma_N$ as the observations.
We have tested that this modification of the fBm does not impact on its power spectrum\footnote{
From our tests on cubes with $0.1<H/L<10$ and a spectral index $\gamma=-3.4$ we 
did not observe significant changes of the power spectrum after the modification of the fBm using Eq.~\ref{eq_modif_fbm}.}.
An example of a simulated 3D density field is shown in Fig.~\ref{fig_cubesimu}.

This method is very similar to the exponentiation method described by \cite{elmegreen2002} and \cite{brunt2002}
but it seems closer to the statistical properties of interstellar emission.
At a given scale, our method produces a fluctuations distribution with a larger FWHM and less
extreme values (lower skewness and kurtosis) 
which is closer to the brightness fluctuations of dust emission at 100~$\mu$m
for instance \cite[]{miville-deschenes2007}.

}

\subsection{Physical properties}

For each normalized simulation we obtain directly a set of quantities that are independent of the
distance $D$ to the cloud: the density contrast ($\sigma_n/<n>$)\footnote{Both the average
density $<n>$ and the standard deviation of the density $\sigma_n$ depend on $D$ but
not their ratio}, 
the standard deviation of the velocity field ($\sigma_v$) and the turbulent
broadening on each line of sight ($\sigma_{turb}(x,y)$ using Eq.~\ref{eq_sigmaturb}). 
By comparing the average $<\sigma_{turb}^2>$ to the observed averaged line width $<\sigma^2>$
(both quantities being averaged over the field) one can estimate the average gas temperature
\begin{equation}
<T> = \frac{m}{k}\left( <\sigma>^2 - \sigma_{psf}^2 - <\sigma_{turb}>^2 \right). 
\end{equation}
and the average Mach number at the scale of the whole cloud 
\begin{equation}
<M>=\sigma_v\sqrt{\frac{m}{k<T>}}
\end{equation}
where $\sigma_v$ is the standard deviation of the 3D velocity field $v$. These are also independent
of $D$.

Assuming a distance $D$ to the cloud, $L=D\tan \theta$ where $\theta$ is the angular size of the GBT field ($1.1^\circ$)
one can also compute the average gas density 
\begin{equation}
<n>=\frac{L}{H}\frac{<N_H>}{D \tan\theta}.
\end{equation}
and the average pressure $<P>=<n><T>$.

\begin{figure}
\includegraphics[width=\linewidth, draft=false]{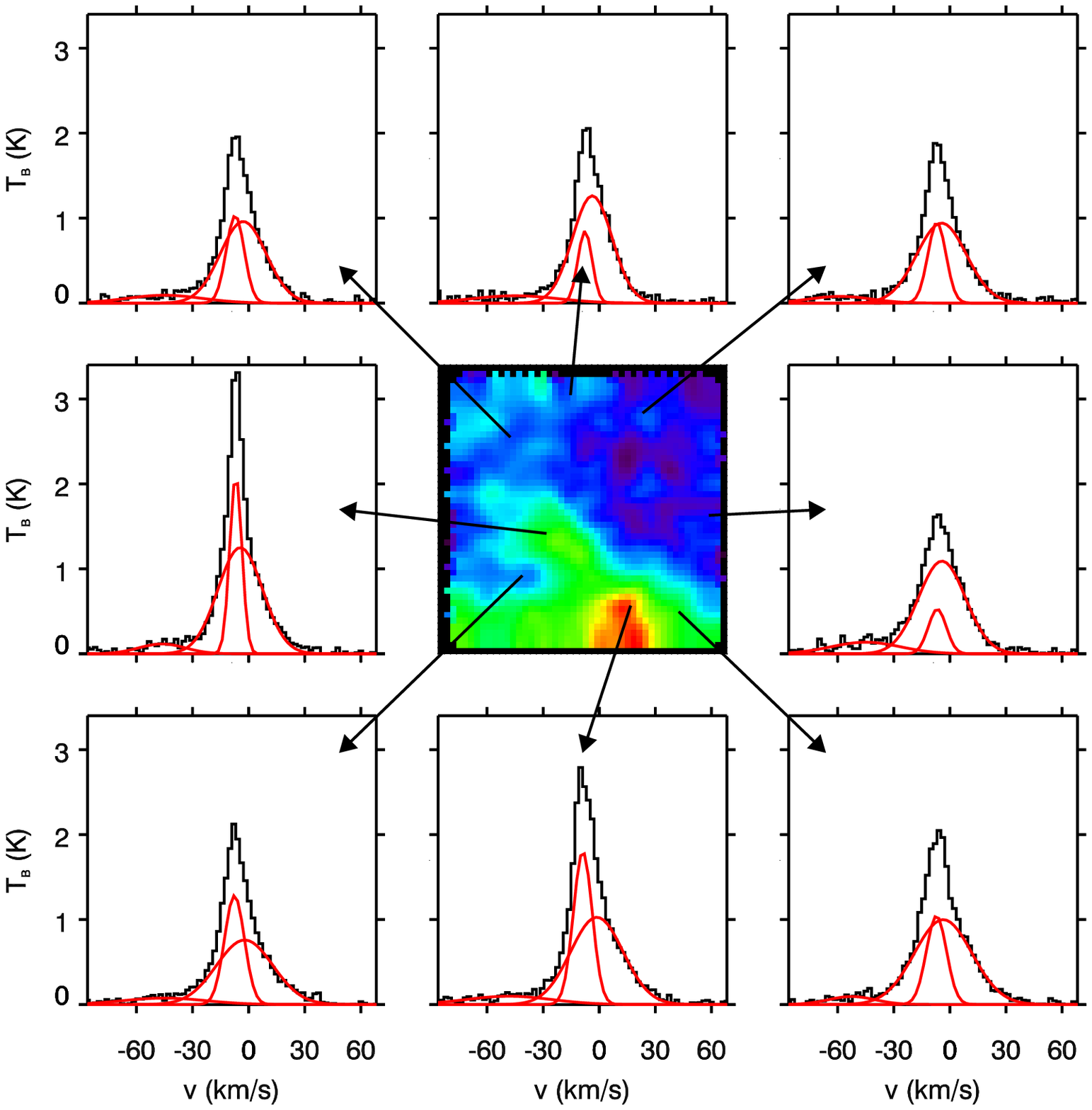}
\caption{\label{fig_exDecomp} Example of the Gaussian decomposition of the 21 cm spectra.
The \hi integrated emission is shown in the central panel, surrounded by 8 typical spectra
(black line) with their associated Gaussian decomposition (red lines). Three components
were used to fit the data : two local components (one narrow and one wide) and 
one intermediate velocity component.}
\end{figure}

\begin{figure*}
\includegraphics[width=\linewidth, draft=false]{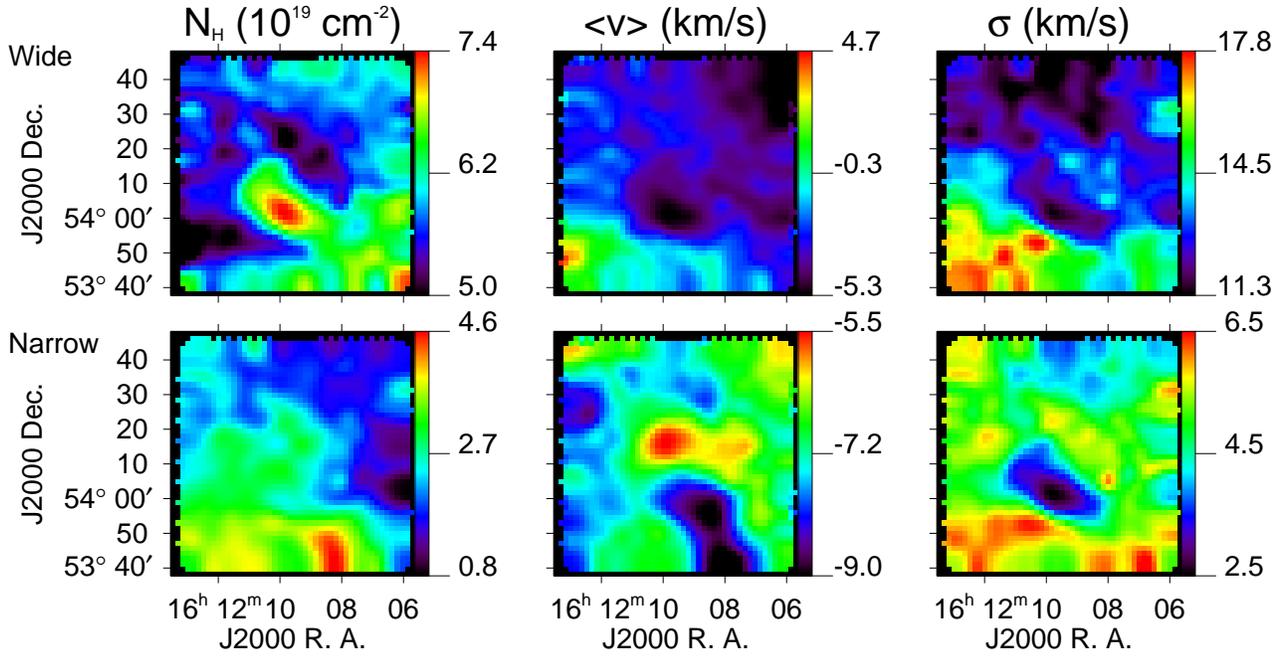}
\caption{\label{fig_resDecomp_TwoCpn} Column density (left column),
centroid velocity (middle column) and velocity dispersion (right column) of the
21 cm emission of the narrow (bottom) and wide (top) components in the local-velocity gas. 
Note the widely different ranges indicated on the color bars.}
\end{figure*}

\subsection{Results}

\label{sec_res_one_component}

The results of these normalized simulations are compiled in Fig.~\ref{fig_resSimulOneCpn}.
On the top four panels are
the density contrast $\sigma_n/<n>$, the turbulent broadening $\sigma_{turb}$, 
the average temperature $<T>$ - in units of 1000~K - and the average Mach number $<M>$,
as a function of the axis ratio $H/L$. The solid line is from the average 
for the 150 realizations done at a given $H/L$ and
the error bars represent the standard deviation.

The bottom two panels of Fig.~\ref{fig_resSimulOneCpn}
give the average density and pressure as a function
of $H/L$ and for three distances $D=75$, 150 and 450~pc. 
In the absence of any measurement of the distance to the cloud we have
estimated a range and a most probable value. 
\cite{lockman1991} showed that the scale height (HWHM) of \hi is 125 pc in the solar neighborhood.
At the Galactic latitude of FN1 ($b=45.1^\circ$) we can consider that most of the clouds lies
within 150~pc ($1\sigma$ of the scale height) from the Sun, and 
that 99\% of the \hi is closer than 450~pc. In addition, using a mapping of NaI absorption
against local stars \cite{sfeir1999} showed that there is very little neutral 
gas at a distance $<75$~pc from the Sun. 
So we consider 75 and 450~pc as the lower and upper limit on distance and 150~pc as the most
probable value.

As expected the required density contrast and the turbulent broadening increases
with $H/L$. The brightness fluctuations average out as the depth of the line
of sight increases, and so stronger density contrasts are needed 
to explain the observed level of column density fluctuations ($\sigma_N$).
The same is true for the turbulent broadening: for deeper clouds
the central limit theorem tends to wash out the differences of centroid velocity
between two adjacent lines of sight, and therefore, to explain a given
level of fluctuations in the centroid velocity map ($\sigma_C$), one needs
stronger turbulent motions.

\subsection{Line profiles}

Another important way of analyzing the results of these simulations is to inspect the shape of
simulated 21 cm spectra. In Fig.~\ref{fig_simu_one_component} we show typical
simulated 21 cm spectra, averaged over the $1.1^\circ \times 1.1^\circ$ field, 
for different values of $H/L$. On the top row the spectra represent what would be
observed with no thermal and instrumental broadenings: the spectral structure is only due to
turbulent motions. As discussed, with the increase of the depth of the cloud
the spectra get wider but also smoother due to the averaging of density and velocity
fluctuations on the line of sight.
Compared to the GBT observations in the bottom row, this turbulent line profile is very narrow.
For this specific field, we have to reject the hypothesis
that the observed spectra is a product of highly turbulent cold gas: the fluctuations seen
in the centroid velocity maps are to small for that.

On the bottom row, we show the simulated spectra when thermal and instrumental 
broadening are added (solid line)
to reproduce the observed line width; these curves have the same velocity dispersion, 
as defined by Eq.~\ref{eq_dispersion1}, as the average GBT spectrum (diamond).
The most important result here is the fact that the simulated spectra do not
reproduce well the observations. The simulated spectra in the bottom row are very close to Gaussian:
the turbulent contribution being relatively small, strong thermal broadening ($T=13000-15000$~K) 
is needed to get to the $\sigma$ observed. More specifically the simulated spectra
failed to reproduce the peaked central part and the extended wings of the observed spectrum.
The conclusion drawn from this analysis is that gas at a single hot temperature
cannot reproduce the observed spectra adequately either. 
The fact that the observed spectra have a relatively narrow core and extended wings
is instead suggestive of two components, one colder than the other.
In the following we explore this possibility.

\begin{figure*}
\includegraphics[width=\linewidth, draft=false]{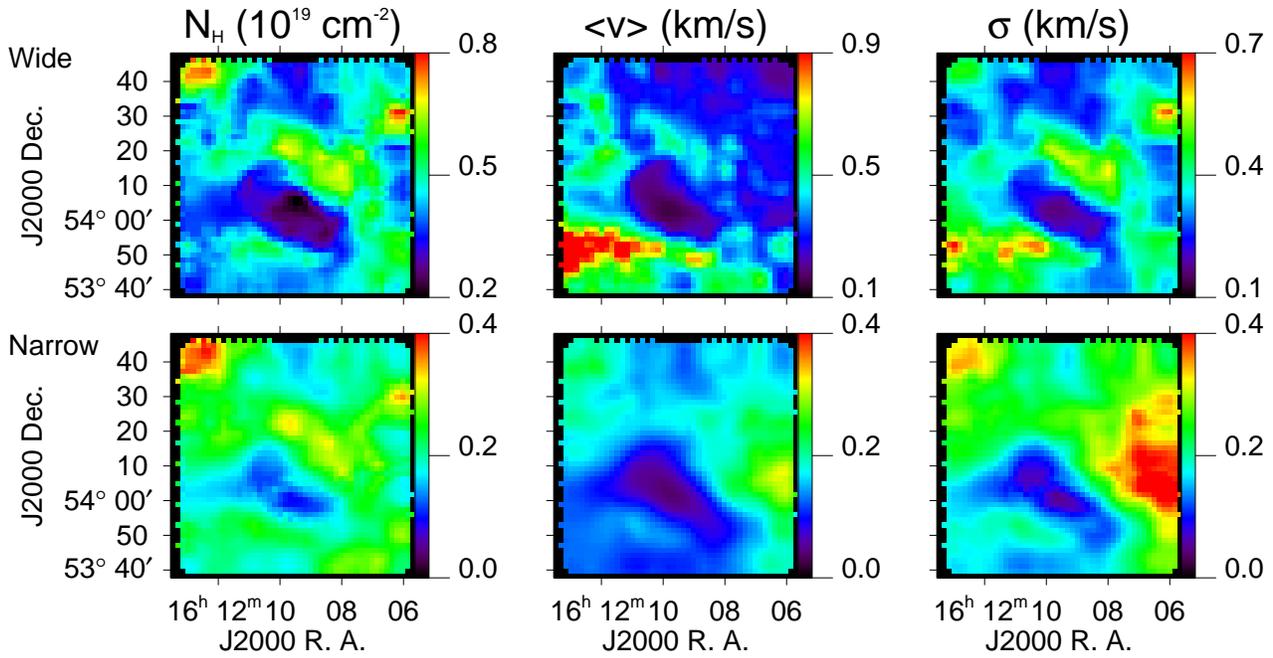}
\caption{\label{fig_error_TwoCpn} Uncertainty maps of the maps shown in Fig.~\ref{fig_resDecomp_TwoCpn}.}
\end{figure*}

\begin{figure*}
\includegraphics[width=\linewidth, draft=false]{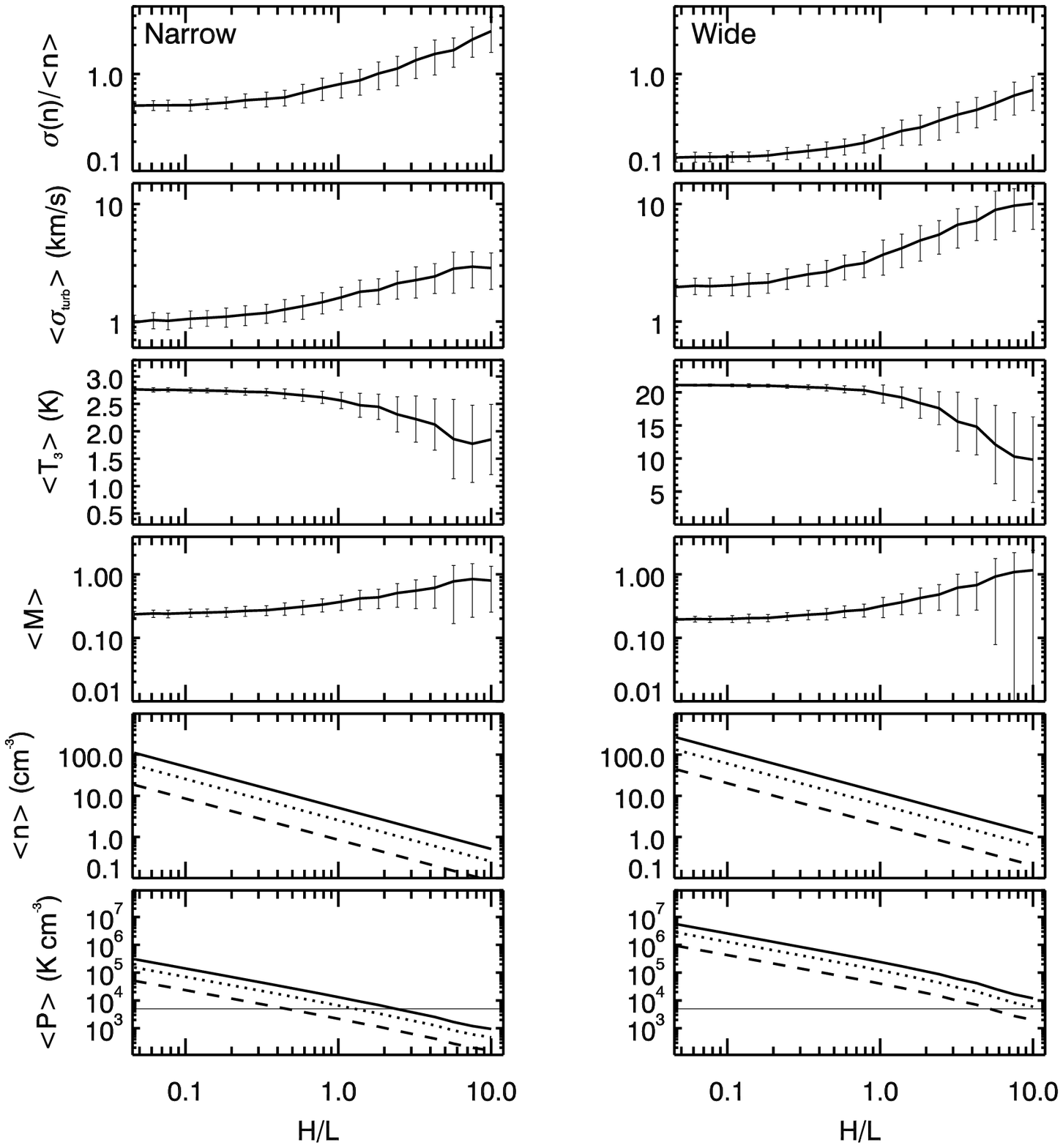}
\caption{\label{fig_resSimulTwoCpn}  Two components model : physical parameters (density contrast $\sigma(n)/<n>$, 
average turbulent broadening $<\sigma_{turb}>$, average temperature $<T_3>=T/1000$, average Mach number $<M>$)
as a function of the axis ratio $H/L$. Are also given the average density $<n>$ and average pressure $<P>$
for distances to the cloud of 75 pc (solid), 150 pc (dotted) and 450 pc (dashed). }
\end{figure*}

\section{Two components model}

\label{sec_two_components}

\subsection{Gaussian decomposition}

\label{sec_gaussian_decomposition}

Many studies of 21 cm observations rely on a Gaussian decomposition of the spectra and make the
assumption that each Gaussian component is a physical structure (a cloud) 
on the line of sight (e.g., \cite{heiles2003a}).
On the other hand, it is not always conceptually easy to reconcile such a decomposition
with the fact that the interstellar medium
has structure at all scales. In fact the Gaussian decomposition is 
facilitated by the significant thermal broadening of the 21 cm line
and by the instrumental response (spectral and spatial) which often have quasi-Gaussian
functions. In addition, with an increase of the path length along a line of sight, the central limit
theorem also tends to bring the velocity and density fluctuations closer to a Gaussian distribution.

As stated by \cite{heiles2003} the Gaussian fitting procedure is also subjective and non-unique.
To make a Gaussian decomposition of the data we used a constrained fit method
that 1) limits the parameter space (centroid and sigma) 
for each component and 2) iterates to find a spatially smooth solution for $N$, $<v>$ and $\sigma$,
which reduces slightly the subjectivity of the fit and certainly increases its robustness.

The spectra are in fact very well fitted by a sum of three Gaussian components, 
a wide one typical of the Warm Neutral Medium (WNM),
a narrow one representing colder gas and a very weak one at intermediate velocity.
Typical fits are preseted in Fig.~\ref{fig_exDecomp}.
The intermediate velocity component is very weak (average column density is $0.8\times 10^{19}$~\ccap, 
maximum is $1.3\times 10^{19}$~\ccap). It probably does represent a physically
distinct component and will not be used in the following analysis.
The resulting column density, velocity centroid and velocity dispersion found
for the two local components are shown in Fig.~\ref{fig_resDecomp_TwoCpn}. 
The uncertainty maps computed using the uncertainties on the fit parameters are 
shown for each quantity in Fig.~\ref{fig_error_TwoCpn}. 
The statistical properties of the integrated emission, centroid
velocity and velocity dispersion of the narrow and wide components are given in Table~\ref{table_gbt}.
The uncertainty on these statistical properties were computed using 1000 realizations of each parameter 
maps shown in Fig.~\ref{fig_resDecomp_TwoCpn} to which we have added a random value at each position. 
The random value was taken from a Gaussian distribution of width given by the uncertainty on the 
fit parameters given in Fig.~\ref{fig_error_TwoCpn}.
The uncertainties given in Table~\ref{table_gbt} are the variance (for the average) or the 
bias (for the standard deviation) of the resulting statistical properties.

This decomposition shows that the column density of the field is 
dominated by the wide component ($\sim 2/3$). On the other hand 
the fluctuations of the total integrated emission are dominated 
by the narrow component. For the discussion in \S~\ref{section_simulation} 
this is compatible with the wide component having a depth much larger than the narrow component. 
We note a significant increase of the wide component column density at
the tip of the main filament of the field, which does not show up
in the narrow component column density. On the other hand the narrow component 
reaches its smallest velocity dispersion (2.5~\kmsp) at that position
which coincide spatially with an interesting convergence of the centroid velocity of the warm component.
This spatial correlation of the column density of the wide component and the 
velocity dispersion of the narrow component at the tip of the filament 
is significant and can't be explained by an artifact of the fitting procedure. 
As it can be seen in Fig.~\ref{fig_error_TwoCpn}, this is the location
where the constraints on the fit are the strongest.

\subsection{Simulations}

To estimate the physical properties of the narrow and wide components we
used the exact same method as described previously. We made fBm simulations of 3D density
and velocity fields
in order to reproduce the observed statistical properties of the narrow and wide
components, as a function the axis ratio $H/L$.
In the absence of constraints we made the assumption that the narrow and wide 
components are both characterized by the
same power spectrum ($\gamma_n=\gamma_v=-3.4$) and that they are uncorrelated. 
Like for the previous simulations we made 150 realizations per $H/L$ value 
both for the narrow and wide components. 
{We also made 30 realizations for
$\gamma_n=\gamma_v=-3.0$ and for $\gamma_n=\gamma_v=-3.8$ in order to test the
robustness of our results on the uncertainty on the power spectrum slope.}

\subsection{Results}

\begin{figure}
\includegraphics[width=\linewidth, draft=false]{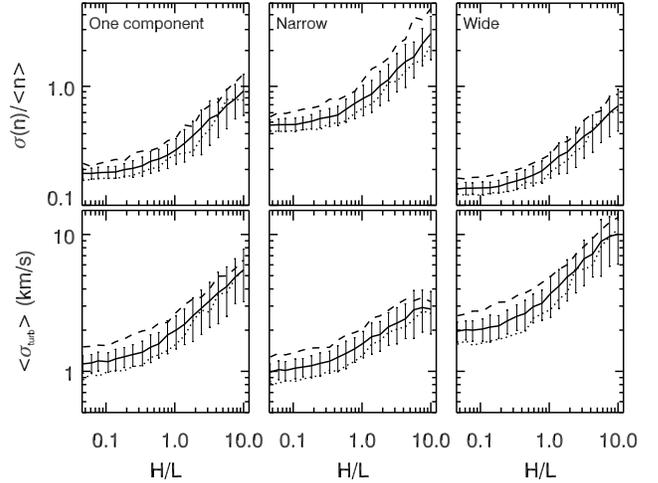}
\caption{\label{fig_testPsdSlope} 
{Effect of a variation of the power spectrum index ($\gamma_n$ and $\gamma_v$) 
for the one component model (left column) and the two components models (middle and right columns).
Result of the simulation for the density contrast $\sigma(n)/<v>$ and turbulent broadening $<\sigma_{turb}>$
when one considers a power spectrum index
of $\gamma_n=\gamma_v=-3.0$ (dashed) and $\gamma_n=\gamma_v=-3.8$ (dotted) (30 realizations per $H/L$ value). 
The solid line with error bars is the results for $\gamma_n=\gamma_v=-3.4$ 
(150 realizations per $H/L$ value - the same curves as shown in figures~\ref{fig_resSimulOneCpn} and \ref{fig_resSimulTwoCpn}). }
}
\end{figure}

The results of the simulations are compiled in 
figures~\ref{fig_resSimulTwoCpn} and \ref{fig_testPsdSlope}. 
Like for the simulations done in the one-component hypothesis, the density contrast and
turbulent broadening increase with $H/L$ but with different average values.
At a given $H/L$ the narrow component has a higher contrast and a lower turbulent velocity dispersion
than the wide component.

{The effect of the uncertainty on the
power spectrum slope on these two quantities is shown in \ref{fig_testPsdSlope}. 
There is a bias introduced by the value of $\gamma_n$ and $\gamma_v$ used. At a given $H/L$, 
steeper (flatter) power spectra leads to lower (larger) density contrast and turbulent broadening.
Nevertheless, even for the extreme values of the power spectrum index used here (-3.0 and -3.8) 
we find very similar results than for  $\gamma_n = \gamma_v =-3.4$.}

\begin{figure*}
\includegraphics[width=\linewidth, draft=false]{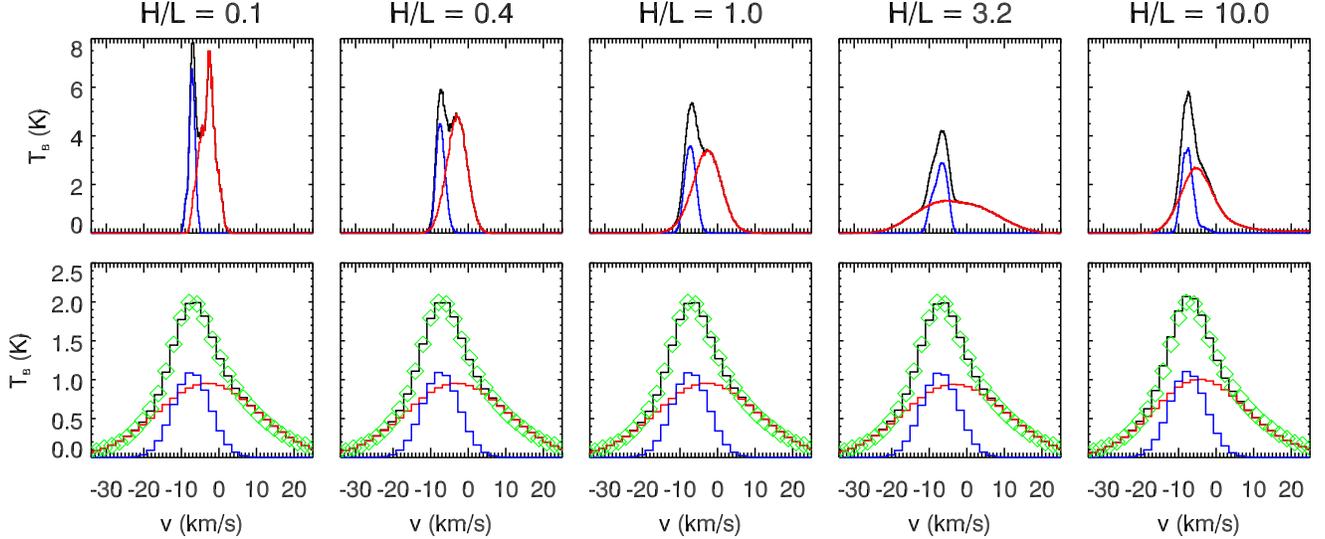}
\caption{\label{fig_simu_two_components} Comparison of simulated and observed spectra.
Same as Fig.~\ref{fig_simu_one_component} but for the narrow (blue curve) and wide (red curve)
components. The black curves on all plots is the sum of the narrow and wide components.}
\end{figure*}

Typical simulated spectra are compared to the average observed spectrum
in Fig.~\ref{fig_simu_two_components}. As in Fig.~\ref{fig_simu_one_component}, 
on that latter figure the top panels show typical
narrow (blue) and wide (red) spectra if there was only turbulence contributing to the broadening. 
On the bottom panels thermal and instrumental
broadening were added to the narrow and wide spectra to match the corresponding 
velocity dispersions $<\sigma>$ given in Table~\ref{table_gbt}. Contrary to the
one component model (see Fig.~\ref{fig_simu_one_component}) the sum of the two components, 
shown in black, matches very well the observations, shown as diamonds.

This simulation, based on a simple representation of the density and
velocity fields (self-similar), gives insight into the physical parameters of the \hi
in this very diffuse region of the Galaxy. If we consider a typical pressure ($P\sim 5000$~K~\ccup)
and distance ($D\sim150$~pc) for this high latitude cloud, the bottom panels
of Fig.~\ref{fig_resSimulTwoCpn} suggests that
the $H/L$ ratio of the narrow component is about 1 but reaches 10 for the wide component. The
wide component is therefore much more spread out on the line of sight than
the narrow one which is more confined spatially. 
In \S~\ref{section_ps_leiden} we noted a break in the Leiden/Dwingeloo data power spectrum 
at a scale $\sim 10^\circ$ (see Fig.~\ref{fig_powerspectra_leiden}), also an indication 
that $H/L\sim10$ for the wide component which dominates the column density.

Using these values of $H/L$ as basis, we found that $\sigma_n/<n> \sim 0.8$ for both
phases, which are relatively low contrast values\footnote{These relatively low contrasts give an indication that 
the velocity power spectrum obtained from the centroid velocity is not affected significantly by density 
fluctuations.}.
These values of density contrasts are compatible with what is usually obtained in MHD
simulations of interstellar clouds \cite[]{brunt2004}. 
We also find that the turbulent broadening is $\sim 2$~km~ s$^{-1}$ for the narrow component
and $\sim 10$~km~ s$^{-1}$ for the wide component. 

In the hypothesis that $H/L=1$ for the narrow component and $H/L=10$ for the wide component, 
the wide component has a temperature ($T\sim 10 000$~K)
and a density ($<n> \sim 0.5-1$) which are typical of WNM gas. 
On the other hand the narrow component has an average temperature ($T\sim 2300-2800$~K)
that is clearly not representative of CNM gas (the median CNM temperature obtained
by \cite{heiles2003} is 70~K). The temperature of the narrow component is
in fact in the thermally unstable range. 
Globally CNM type gas does not contribute a large fraction of the column density in the GBT field.
This is not totally unexpected as \cite{heiles2003} found several lines of sight with no CNM.
This analysis in fact leads to the conclusion that, in this field, $\sim 30\%$
of the total \hi column density is in the thermally unstable regime.
This is similar to the average obtained by \cite{heiles2003} using a completely
different technique. In addition, we argue that the unstable phase has a 
lower filling factor (lower $H/L$) than the WNM. 
In that respect the unstable gas would have a structure closer to the CNM.

Regarding the turbulent motions, the unstable gas is
definitely sub-sonic with $<M>\sim 0.3$. On the other hand the WNM gas
is quite turbulent and likely to be supersonic with $<M> \sim1$. 
This could give an indication on the dissipation of turbulent motions 
associated with the phase transition (WNM$\rightarrow$unstable$\rightarrow$CNM).

Globally our results are in good accord with the simulations by \cite{audit2005} of
the \hi thermal instability. They
show that, for slightly turbulent fields, up to 30\% of the gas is in the
thermally unstable phase ($200<T<5000$~K) with an average density $n\sim 5$~cm$^{-3}$.
For $H/L=1$ and $D\sim 150$~pc we also find $<n>\sim5$~\ccu for our narrow component.
It is also interesting that \cite{audit2005} found the thermally unstable gas to be 
filamentary in accord with the low $H/L$ ratio of the observed narrow component. 

\subsection{Formation of a CNM structure ?}

The analysis above gives average values over the field of view
but one can also estimate the physical parameters of a specific structure
seen in the column density map, like the main filament for instance which has a low velocity
dispersion and could contain classical CNM gas.
This filament has an angular transverse size of 10 arcmin (which is barely resolved by the GBT).

Considering the properties of the narrow component at the position of minimum $\sigma$
($N_H \sim 3.3 \times 10^{19}$~\cca and $\sigma=2.5$~\kmsp) and assuming $H/L=1$
one can derive its depth
\begin{equation}
H = 0.44\, D_{150} \; \mbox{(pc)},
\end{equation}
its average density
\begin{equation}
<n> = 24 / D_{150} \; \mbox{(cm$^{-3}$)}
\end{equation}
average temperature
\begin{equation}
<T> = 204 \, P_{5000} \, D_{150} \; \mbox{(K)}
\end{equation}
turbulent broadening
\begin{equation}
\sigma_{turb} = \sqrt{5.1 - 1.7\, P_{5000} \, D_{150} } \; \mbox{(km/s)}
\end{equation}
and average Mach number
\begin{equation}
<M> = \sqrt{\frac{3}{P_{5000} \, D_{150}} -1}
\end{equation}
where $D_{150}=D/(150$~pc) and $P_{5000}=P/(5000$~K~cm$^{-3}$).
For those typical values of distance and pressure we get $\sigma_{turb}=1.8$~\kms
and $<M>=1.4$.
These physical parameters are typical of CNM gas, 
we therefore conclude that it is likely that a significant fraction of the column density
at the tip of the filament seen in the FN1 field is CNM gas. 
The fact that we observe opposite velocity gradients of WNM gas at the 
same position (see Fig.~\ref{fig_resDecomp_TwoCpn})
is reminiscent of what is seen in the simulations of \cite{audit2005} where 
CNM structures are efficiently formed in a large scale convergent flow. 

\section{Conclusion}
 
\label{sec_conclusion}

In this paper we presented a new technique
to constrain the physical properties of the 3D density and velocity fields of \hi clouds.
We showed that from a set of statistical quantities computed with the observations
($<N_H>$, $\sigma_{N}$, $<C>$, $\sigma_C$, $<\sigma>$, $\gamma_{N}$, 
$\gamma_C$) and the spectral shape of the 21 cm line, it is possible to 
constrain the physical properties of the \hi on the line of sight like
density contrast, turbulent broadening, average temperature and average Mach number.

{The basis of the method is to constrain the underlying physical properties of the cloud by
producing a large number of 3D realizations of the density and velocity fields with proper statistical properties.
This type of method is routinely used in the analysis of the cosmic microwave background (CMB).
One important difference with the CMB analysis is the fact that we do not have a complete
description of the statistical properties of interstellar fields and that Gaussian random
fields (or fBm) are not a good representation of the observations.
One known limitation of fBms for the analysis of interstellar matter 
is the fact that they are isotropic and Gaussian objects.
In this analysis we used a method to modify classical fBm objects such that they
reproduce the power spectrum and the non-Gaussian properties of density fields 
observed in numerical simulations and observations of interstellar clouds.
A further step would be to produce such artificial fields with anisotropic structure
(obvious in interstellar emission) and a better description of the non-Gaussianity
of the velocity field (weaker than the non-Gaussianity of the density field).
Nevertheless we argue that the statistical properties used here to estimate the
3D physical properties of the gas are more controlled by the power-law power spectrum
of the density and velocity fields (and by the non-Gaussian 
properties of the density field) rather than by their anisotropy. 
}

We applied this new technique to analyze 21 cm observations obtained at the GBT
of the FN1 field, a very diffuse region at high Galactic latitude used for extra-galactic studies.
First we showed that the observed spectra of FN1 cannot be explained by CNM type gas affected by 
strong turbulent motions. We were able to exclude this hypothesis by showing that
it is incompatible with the statistical properties of the centroid velocity.
More importantly we showed that the shape of the spectra (narrow core and extended wings)
cannot be well reproduced with only one \hi component, whatever its temperature.

Instead the observations are well fitted by a combination of two spectral components: 
one narrow ($<\sigma>=5.0$~\kmsp) that traces thermally unstable gas, 
and one wide ($<\sigma>=13.4$~\kmsp) which has the physical properties of the WNM.
The results of our analysis show that there is no significant CNM type gas in the field
and that the thermally unstable phase contributes $\sim30$\% of the column density,
near the average reported in recent observational results \cite[]{heiles2003} and numerical simulations
\cite[]{audit2005}.
Both components have a density contrast $\sigma_n/<n>$ close to 0.8 but the WNM component
is much more spread out on the line of sight with a filling factor 10 times higher than
the unstable gas. We also conclude that the WNM gas is mildly supersonic ($<M>\sim1$)
and the unstable phase is definitely sub-sonic ($<M>\sim0.3$).

Finally this portion of the FN1 field contains a narrow filament (axis ratio $>4$), 
almost unresolved by the GBT, which has a low velocity dispersion (2.5~\kmsp).
Our analysis indicates that this structure, which could contain CNM type gas,
is found at the locus of opposite velocity gradients in the WNM
which could have triggered a phase transition.

{\noindent\it Acknowledgments.} The authors thanks Felix Jay Lockman for providing
the GBT observations used in this analysis and the anonymous referee for very constructive
comments.

\end{document}